\documentclass[11pt,a4paper]{article}
\usepackage[hyperref]{acl2020}
\usepackage{times}
\usepackage{latexsym}
\usepackage[pdftex]{graphicx}
\usepackage{amssymb}
\usepackage{amsthm}
\usepackage{amssymb}
\usepackage{amsmath}  

\usepackage{microtype}

\aclfinalcopy 

\title{Amplifying Emotional Signals: Data-Efficient Deep Learning for Robust Speech Emotion Recognition}

\author{Tai Vu \\
  Department of Computer Science \\
  Stanford University \\
  \texttt{taivu@stanford.edu} 
}

\date{}

\begin{document}
\maketitle

\begin{abstract}
Speech Emotion Recognition (SER) presents a significant yet persistent challenge in human-computer interaction. While deep learning has advanced spoken language processing, achieving high performance on limited datasets remains a critical hurdle. This paper confronts this issue by developing and evaluating a suite of machine learning models, including Support Vector Machines (SVMs), Long Short-Term Memory networks (LSTMs), and Convolutional Neural Networks (CNNs), for automated emotion classification in human speech. We demonstrate that by strategically employing transfer learning and innovative data augmentation techniques, our models can achieve impressive performance despite the constraints of a relatively small dataset. Our most effective model, a ResNet34 architecture, establishes a new performance benchmark on the combined RAVDESS and SAVEE datasets, attaining an accuracy of $66.7\%$ and an F1 score of $0.631$. These results underscore the substantial benefits of leveraging pre-trained models and data augmentation to overcome data scarcity, thereby paving the way for more robust and generalizable SER systems.

\end{abstract}

\section{Introduction}
The ability to perceive and interpret human emotion is a cornerstone of intelligent interaction, yet it remains a significant frontier in the quest to build truly symbiotic human-computer interfaces. While deep learning has catalyzed remarkable progress in spoken language processing \citep{lecun1995convolutional, hochreiter1997long, vaswani2017attention}, LSTMs, particularly in speech recognition and synthesis, the domain of Speech Emotion Recognition (SER) presents a more nuanced and formidable challenge. Unlike transcribing words, recognizing emotion requires decoding subtle, complex acoustic cues, variations in pitch, tone, and energy, that are often ambiguous and highly context-dependent. This complexity is magnified by a persistent bottleneck in the field: the scarcity of large-scale, emotionally annotated speech datasets. Consequently, even sophisticated neural network architectures frequently struggle to generalize from limited training data, leading to overfitting and unreliable performance in real-world applications.

This paper directly confronts the critical challenge of data scarcity in SER. We propose and validate a methodology that demonstrates how state-of-the-art performance can be achieved even when training data is limited. By systematically evaluating a range of models, from traditional Support Vector Machines (SVMs) to deep learning architectures like LSTMs and CNNs, we identify the most effective approaches for this constrained environment. Our core contribution lies in the strategic application of transfer learning and data augmentation. We treat log-mel spectrograms, 2D visual representations of audio, as images, allowing us to leverage the power of a ResNet34 model pre-trained on the vast ImageNet database. This cross-domain knowledge transfer, combined with a suite of augmentation techniques, empowers our model to learn robust, generalizable features from a small dataset. Our experiments, conducted on a composite of the RAVDESS and SAVEE datasets, culminate in a model that not only achieves a new benchmark with $66.7\%$ accuracy and a $0.631$ F1 score but also provides a clear blueprint for developing powerful, data-efficient SER systems.




\section{Related Works}
Research in Speech Emotion Recognition (SER) has evolved significantly, transitioning from traditional machine learning techniques to sophisticated deep learning architectures. This evolution informs our work, which focuses on achieving robust performance in data-constrained, audio-only environments.

\subsection{Traditional and Feature-Based Approaches}
Early explorations into SER relied on engineered acoustic features to classify emotions. For instance, Hidden Markov Models (HMMs) were employed to extract and model features from speech signals for emotion detection \cite{schuller2003hidden}. A pivotal advancement was the adoption of Mel Frequency Cepstral Coefficients (MFCCs), which proved highly effective for speech-related tasks. \citet{demircan2018application} successfully utilized MFCCs extracted from the EMO-DB dataset, combining them with fuzzy C-means clustering and k-nearest neighbors (kNN) for emotion prediction. While foundational, these methods often depend on handcrafted features that may not capture the full complexity of emotional expression in speech.


\subsection{Deep Learning in Audio-Only SER}
With recent breakthroughs in deep learning, focus has shifted toward neural networks that can learn hierarchical features directly from audio representations. \citet{lim2016speech} demonstrated significant improvements over traditional methods by applying Convolutional Neural Network (CNN) and Long Short-Term Memory (LSTM) layers to Short-Time Fourier Transform representations of raw audio. However, a primary limitation of most deep learning systems is their reliance on large quantities of training data to achieve high performance. Our project distinguishes itself by tackling this very issue, demonstrating that the strategic implementation of data augmentation and transfer learning can enable robust model training on a relatively small database, overcoming common issues of data scarcity and overfitting.


\subsection{Multimodal Emotion Recognition}
A parallel stream of research has focused on enhancing SER by incorporating information from additional modalities, such as video or text. For example, some studies have combined hand-crafted speech features like pitch and energy with facial landmark features extracted from videos \citet{kim2013deep}. \citet{tzirakis2017end} developed a more advanced multimodal system using 1D convolutional layers for speech and a ResNet50 for visual information from video frames, feeding the combined features into an LSTM for final classification. While these multimodal approaches have shown improved accuracy, they are contingent on the availability of visual or textual data, which is often not the case. This limitation underscores the importance of developing high-performing, audio-only systems. Our work is therefore motivated by the need for a practical solution that can accurately infer emotion based solely on audio inputs, without reliance on supplementary data streams.



\section{Approach}

\subsection{Models}

Our machine learning system included an encoder, which was followed by a classifier. The encoder received an audio clip and then produced a vector representation of the input data. Subsequently, this encoding was fed into the classifier, which outputted an emotion label.

\subsubsection*{Model 1: MFCC and SVM}

As a starting point, we implemented the feature extractor using the Mel
Frequency Cepstral Coefficients (MFCC). Afterwards, we took the averages of these MFCC input features across the time dimension and then used them to train a Support Vector Machines (SVM) model \citep{bosertraining} to classify different emotions.

\subsubsection*{Model 2: Log mel spectrograms and LSTM}

Our second model encoded each data point by computing a mel-scaled spectrogram and then converting it to log space. We built an LSTM neural network as our classifier. This network contained $2$ bidirectional LSTM layers, followed by a dropout layer, a linear layer, and a softmax layer.

\subsubsection*{Model 3: Log mel spectrograms and CNN}

\begin{figure}
    \centering
    \includegraphics[width=0.45\textwidth]{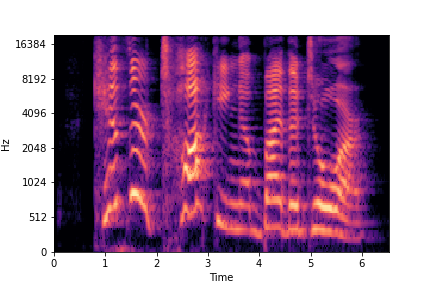}
    \caption{Log mel spectrogram features of an example.}
    \label{fig:logmel}
\end{figure}

In this model, we also extracted log-scaled mel spectrograms for the input speech data. Since these features were similar to 2D image arrays (shown in Figure \ref{fig:logmel}), we then fed them into a CNN classifier in order to obtain emotion labels. Previously, we intended to put raw waveforms directly through the CNN model. However, during our experiments, we found out that training the CNN on log-scaled mel spectrograms was easier and more stable.

We chose ResNet34 \citep{he2016deep} as our CNN architecture. Additionally, we experimented with two different approaches: training a ResNet34 network from scratch and using transfer learning to finetune a ResNet34 model that was pretrained on the ImageNet database \citep{russakovsky2015imagenet, vu2020flopcombiningregularizationpruning}.

\subsection{Data Augmentation}

As we developed and trained our models on a small speech dataset, data augmentation would be helpful in generating more training data and dealing with overfitting problems.

\subsubsection*{Image-based Data Augmentation}

In particular, since our CNN models were trained on image-like 2D arrays of log-scaled mel spectrograms, we applied several data augmentation methods on these input data, which include rotating by a small degree, zooming in, and changing brightness. Although such image-based augmentation techniques were more common in computer vision tasks and were not directly applied to audio data, we will demonstrate in Section 5.4 that these techniques indeed helped prevent overfitting and improve model performance.

\subsubsection*{Progressive Resizing}

Another augmentation method that we used was progressive resizing \citep{colangelo2021progressive, vu2025bertvqavisualquestionanswering}. Specifically, we first trained the CNN models on smaller versions of the log-scaled mel spectrogram arrays $(128 \times 128)$, and then finetuned the networks on arrays of larger sizes $(256 \times 256)$. This approach not only augmented the training data, but also allowed the models to train much faster.

\subsubsection*{Mixup}

In addition, we harnessed Mixup, a data augmentation technique that generated convex combinations of pairs of training examples and their labels \citep{zhang2018mixup}. Particularly, for two randomly sampled data points $(x_i, y_i)$ and $(x_j, y_j)$, this method constructed a new example of the form:
\begin{equation*}
\begin{aligned}
\tilde x &=& \lambda x_i + (1 - \lambda) x_j \\
\tilde y &=& \lambda y_i + (1 - \lambda) y_j
\end{aligned} 
\end{equation*}

Here, $x_i, x_j$ are input vectors, $y_i, y_j$ are one-hot label encodings, and $\lambda \in [0, 1]$. In this way, Mixup acted as a regularizer that encouraged the linear behaviors of the models, reduced their variance, and enhanced their generalization powers.

\section{Implementation}

I implemented the code for this project in Python using PyTorch \citep{paszke2019pytorch}, FastAI \citep{howard2020fastai}, Scikit-learn \citep{pedregosa2011scikit}, Librosa \citep{mcfee2015librosa}. All the code can be found \href{https://github.com/taivu1998/ML-SER}{here}.

\section{Experiments}

\subsection{Data}

In this project, we used the Ryerson Audio-Visual Database of Emotional Speech and Song (RAVDESS) database \citep{livingstone_steven_r_2018_1188976} and the Surrey Audio-Visual Expressed Emotion (SAVEE) database \citep{jackson2014surrey}. We combined them into a single dataset for training and testing our models.

RAVDESS is an English language database that contains $1440$ utterances. This dataset was made by $24$ actors ($12$ female and $12$ male), who said two sentences "Kids are talking by the door" and "Dogs are sitting by the door" with various emotions. Meanwhile, the SAVEE database consists of $480$ audio clips created by $4$ male actors, and each of them recorded $15$ sentences. There are $8$ different emotion classes, including \texttt{neutral}, \texttt{calm}, \texttt{happy}, \texttt{sad}, \texttt{angry}, \texttt{fearful}, \texttt{disgust}, and \texttt{surprised}.

\begin{figure}
    \centering
    \includegraphics[width=0.5\textwidth]{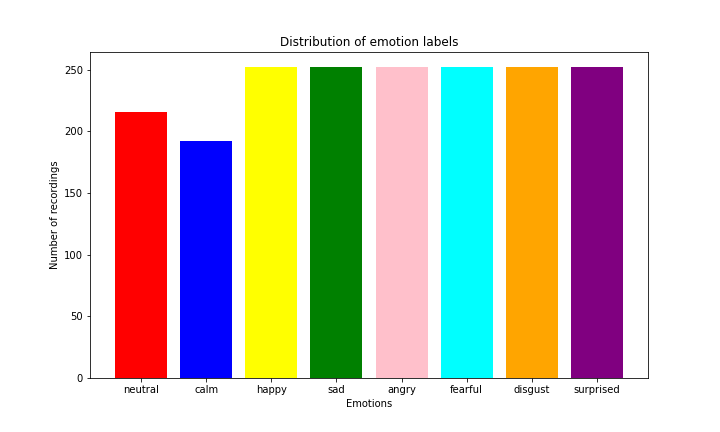}
    \caption{Distribution of emotion labels in the dataset.}
    \label{fig:emotion_dist}
\end{figure}

The duration of each utterance ranges from $3$ to $5$ seconds. The total duration of audio recordings is roughly $2$ hours. In addition, we can see in Figure \ref{fig:emotion_dist} that most of the emotional classes are relatively well balanced. The \texttt{neutral} and \texttt{calm} labels contain slightly fewer audio clips than the other $6$ classes.


We split the dataset into $90\%$ for training, $5\%$ for validation, and $5\%$ for testing. 


\subsection{Experiment Details}

For the SVM model, we produced $20$ MFCCs for each input audio clip. We chose an RBF kernel for the SVM algorithm.

For the LSTM and CNN models, we generated $128$ mel bands when converting input speeches to mel spectrograms. We trained the LSTM model and the vanilla CNN model (with no pretraining) for $200$ epochs. Meanwhile, for the ResNet34 model that was pretrained on ImageNet, we finetuned its weights for $30$ epochs. We used a batch size of $64$ and a learning rate of $0.001$, with a decay rate of $0.9$. We trained the above neural networks using the Cross Entropy loss and the Adam optimization algorithm \citep{kingma2014adam}.

\subsection{Evaluation Methods}

Since this project tackled a classification problem, we used classification accuracy scores and F1 scores for evaluating model performance.

\subsection{Results}

\begin{table*}
\centering
\begin{tabular}{lll}
\hline
\textbf{Models} & \textbf{Accuracy} & \textbf{F1 Scores}\\
\hline
SVM & 51.7\% & 0.509 \\
LSTM & 52.8\% & 0.497 \\
CNN (trained from scratch) & 45.8\% & 0.426 \\
CNN (transfer learning) & 57.3\% & 0.528 \\
CNN (transfer learning, data augmentation) & \textbf{66.7\%} & \textbf{0.631} \\
\hline
\end{tabular}
\caption{Performance of different models on the validation set.}
\label{results}
\end{table*}

As shown in Table \ref{results}, the SVM algorithm produced an accuracy of $51.7\%$ and an F1 score of $0.509$. This result was better than we expected, because the model only took into account the mean values of the MFCC features across the time dimension. In other words, the SVM algorithm did not get access to useful temporal dependencies amongst the input MFCC features, but still learned to predict emotions with more than $50\%$ accuracy.

After that, the LSTM model performed slightly better than the SVM algorithm, with a higher accuracy of $52.8\%$ and a comparable F1 score of $0.497$. When investigating its training process, we can see that the performance was still quite low because the LSTM network was overfitting to the training data. In particular, the model learned to decrease training losses to a small value (around 0.5), but the validation losses were still high (around 2.9).

A similar pattern occurred for the vanilla CNN model (with no pretraining), as it only produced $45.8\%$ accuracy. In this case, another issue is that because the training set was too small, the Resnet34 network was not able to learn good representations of the speech contents, so it could not generalize well to unseen data.

In fact, when we finetuned the ResNet34 model with pretrained weights from ImageNet, the performance went up significantly ($57.3\%$ in accuracy and $0.528$ in F1 score). Therefore, we can see that the neural network learned useful feature representations of the speech data after being pretrained on a large database like ImageNet. When it was finetuned on our small dataset, the model was able to transfer its prior knowledge about images to reading and extracting information from image-like log-scaled mel spectrogram arrays. The finetuning process then helped the model to adapt to the domain of our dataset even better, which enhanced its performance.

\begin{figure}
    \centering
    \includegraphics[width=0.4\textwidth]{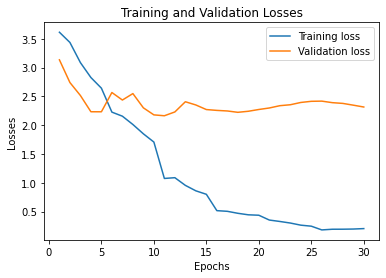}
    \includegraphics[width=0.4\textwidth]{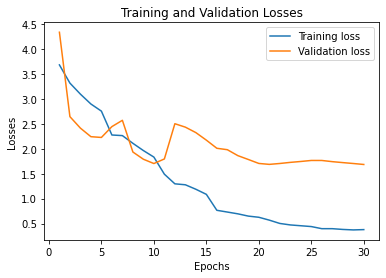}
    \caption{Training and validation losses across 30 epochs for the ResNet34 model without data augmentation (upper) and with data augmentation (lower).}
    \label{fig:cnn_loss}
\end{figure}

Finally, the ResNet34 model with both transfer learning and data augmentation achieved the best performance, with an accuracy of $66.7\%$ and an F1 score of $0.631$. This illustrates the effectiveness of data augmentation techniques in boosting our model performance. Indeed, as we can see in the upper plot of Figure \ref{fig:cnn_loss}, the ResNet34 network without data augmentation was still overfitting, with low training loss values and high validation loss values. This means that the gap between the training losses and the validation losses was still very large. However, this problem was alleviated with the support of data augmentation, as shown in the lower plot of Figure \ref{fig:cnn_loss}. Both the training losses and the validation losses decreased gradually, and the gap between them was significantly narrowed.



\begin{figure}
    \centering
    \includegraphics[width=0.35\textwidth]{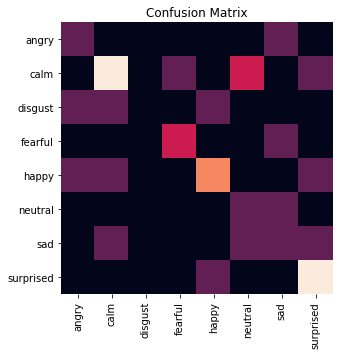}
    \caption{Confusion matrix for the best CNN model.}
    \label{fig:confusion_matrix}
\end{figure}

\begin{figure}
    \centering
    \includegraphics[width=0.45\textwidth]{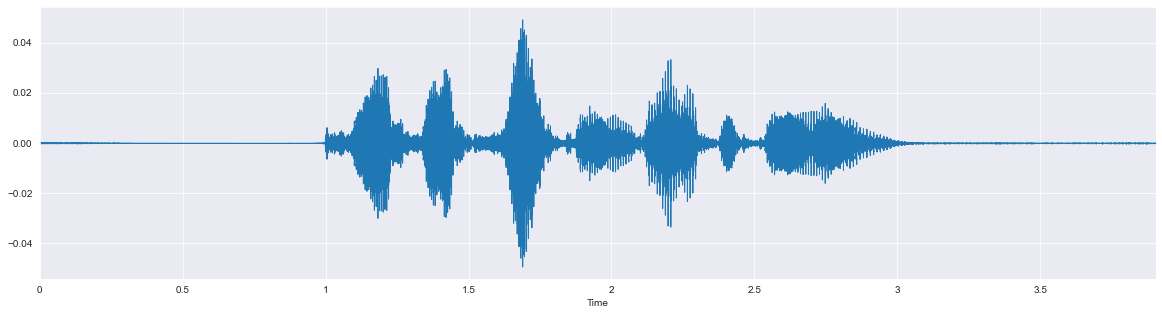}
    \includegraphics[width=0.45\textwidth]{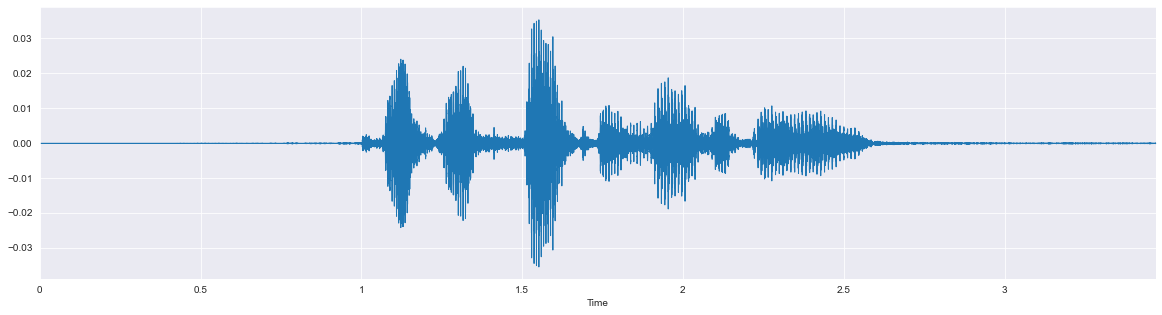}
    \caption{The waveforms of a neutral utterance (upper) and a calm utterance (lower) from the same actor.}
    \label{fig:neutral_calm}
\end{figure}

Meanwhile, because the accuracy of our final model was less than $70\%$, there is still a lot of room for improvement. One of the main challenges faced by our models was that RAVDESS and SAVEE were two simulated datasets, which consisted of several actors repeating the same sentences with various emotions. Hence, the speech contents in these datasets were not diverse enough for our machine learning programs to learn proper representations of input audio data and detect correlations between human speeches and emotions. In addition, we can observe in Figure \ref{fig:confusion_matrix} that the ResNet34 model performed well on certain positive classes like \texttt{surprised}, \texttt{happy}, and \texttt{calm}, while produced lower accuracy on some other negative classes like \texttt{disgust} and \texttt{angry}. Furthermore, there was some confusion between certain pairs of emotion labels, such as \texttt{neutral} and \texttt{calm}. This issue is understandable, because the audio clips from these two classes in our dataset often sound similar. Two examples from those two classes are shown in Figure \ref{fig:neutral_calm}.


\section{Conclusion}
In this study, we successfully addressed the critical challenge of developing a high-performing Speech Emotion Recognition (SER) system in a data-constrained environment. We systematically developed and evaluated a series of models, including SVMs, LSTMs, and CNNs, culminating in a ResNet34 architecture that achieved a promising accuracy of $66.7\%$ and an F1 score of $0.631$ on a combined dataset of RAVDESS and SAVEE utterances.

The key to this achievement was a strategic methodology rooted in two powerful techniques. First, transfer learning proved instrumental in overcoming the limitations of our small dataset. By leveraging a ResNet34 model pre-trained on ImageNet, we effectively transferred rich, hierarchical feature-extraction capabilities to the domain of audio-spectrogram analysis, providing a robust foundation for learning. Second, data augmentation was critical for improving generalization and mitigating overfitting—a significant issue observed in our initial deep learning models. The combination of these methods demonstrates that it is possible to build effective SER systems without relying on massive, annotated speech corpora. While acknowledging the limitations inherent in using simulated datasets and the potential for further accuracy improvements, our work provides a clear and effective blueprint for data-efficient learning in the field of SER.

\section{Future Work}
Building on the promising results of this study, we have identified several key directions for future research that aim to further advance the capabilities and robustness of our SER models.

\subsection{Model and Training Refinements}
Our immediate next step involves more exhaustive hyperparameter tuning to optimize the current ResNet34 architecture. Beyond this, we plan to explore hybrid network architectures, such as combining CNN layers for efficient feature extraction from spectrograms with LSTM layers to better model the temporal dependencies inherent in speech. Such models may offer a more holistic understanding of the acoustic and sequential nature of emotional expression.

\subsection{Advanced Audio-Based Augmentation}
While our image-based augmentation on spectrograms was effective, we plan to implement more domain-specific, audio-based data augmentation techniques. Methods such as pitch shifting, time-stretching, and altering loudness can create more realistic variations in the training data, better preparing the model for real-world acoustic conditions. Furthermore, we intend to integrate advanced techniques like SpecAugment, which has shown exceptional results in speech recognition by masking frequency channels and time steps directly on the spectrogram, thereby forcing the model to learn more robust and redundant features \citep{Park_2019, Vu2025GANimeGA}.

\subsection{Exploring State-of-the-Art Pre-trained Speech Models}
Finally, recognizing the immense potential of transfer learning, our most ambitious future work will involve fine-tuning large-scale models that have been pre-trained specifically on vast unlabeled speech corpora. Models like wav2vec \citep{schneider2019wav2vec} and SpeechBERT \citep{chuang2019speechbert, sun2025privacypreservinginferencepersonalized} have learned fundamental representations of human speech that are potentially far more powerful than those learned from general-purpose image datasets. Adapting these state-of-the-art speech foundation models to the downstream task of emotion recognition could unlock significant performance gains and push the boundaries of what is achievable in SER.

\newpage
\bibliography{anthology,acl2020}
\bibliographystyle{acl_natbib}





\end{document}